# The origin of the ferroelectric-like orthorhombic phase in oxygen-deficient HfO$_{2-y}$ nanoparticles

Eugene A. Eliseev,[1*] Iryna V. Kondakova,[1] Yuri O. Zagorodniy,[1] Hanna V. Shevilakova,[2] Oksana V. Leshchenko,[1] Victor N. Pavlikov,[1] Lesya P. Yurchenko,[1] Myroslav V. Karpets,[1] and Anna N. Morozovska [3†]

[1] Frantsevich Institute for Problems in Materials Science, National Academy of Sciences of Ukraine Kyiv, Ukraine

[2] Department of Microelectronics Igor Sikorsky Kyiv Polytechnic Institute Kyiv, Ukraine

[3] Institute of Physics National Academy of Sciences of Ukraine Kyiv, Ukraine

## Abstract

In this work we established the relationship between the crystalline structure symmetry, point defects and possible appearance of the ferroelectric-like polarization in HfO$_{2-y}$ nanoparticles. Notably, that XRD and EPR analysis revealed the formation of the ferroelectric-like orthorhombic phase in the oxygen-deficient HfO$_{2-y}$ nanoparticles (pure and doped with rare-earth element yttrium). DFT calculations showed that small HfO$_2$ nanoparticles may become polar, especially in the presence of impurity atoms and/or oxygen vacancies. To explain the experimental results, we have modified the effective LGD model through the parameterization approach, focusing on the Landau expansion coefficients associated with the polar (FE) and antipolar (AFE) orderings, which agrees with the performed DFT calculations. The effective LGD model can be useful for the development of the novel generation of silicon-compatible ferroelectric nanomaterials based on the Hf$_x$Zr$_{1-x}$O$_{2-y}$.

[*] Corresponding author: eugene.a.eliseev@gmail.com
[†] Corresponding author: anna.n.morozovska@gmail.com

**INTRODUCTION**

The discovery of ferroelectric and antiferroelectric properties in thin films of lead-free hafnium ($HfO_2$) and zirconium ($ZrO_2$) oxides open them for next-generation of Si-compatible ferroelectric memory elements like ferroelectric random-access memories (FeRAMs) and field-effect transistors (FETs) [1, 2]. However, a comprehensive understanding of the crucial interplay between stoichiometry, oxygen vacancies, size, and surface effects in nanosized $Hf_xZr_{1-x}O_{2-y}$ remains elusive still [3].

Bulk $HfO_2$ and $ZrO_2$, characterized as high-k dielectrics, lack inherent ferroelectric properties across a wide range of standard temperatures and pressures [4, 5]. This behavior undergoes a dramatic shift when these materials are scaled down to the nanoscale, where complex interactions and structural transitions come into play. One key factor influencing the observed ferroelectric properties in $Hf_xZr_{1-x}O_2$ films is the presence of the polar orthorhombic phase. However, this phase suffers from metastability compared to the bulk monoclinic phase, leading to challenges with ferroelectric phase stability. Additionally, the properties of these thin films are highly sensitive to various factors, including substrate material, annealing conditions, deposition method, film thickness, and dopant concentrations [6, 7, 8]. Depending on the interplay of these factors, the value of x in $Hf_{1-x}Zr_xO_2$ films can range from 0 to 1, ultimately dictating whether the material exhibits dielectric, ferroelectric, or antiferroelectric behavior [9, 10].

Extensive theoretical [11, 12, 13] and experimental [14, 15, 16] theoretical and experimental evidence, as highlighted in our previous studies [29, 30, 31], underscores the critical role of surface and grain boundary energy, alongside oxygen vacancies, in optimizing these materials for practical applications in advanced FeRAM and FET technologies. Further research and optimization efforts are still required to fully unlock the full potential of binary oxide $Hf_xZr_{1-x}O_{2-y}$ nanoparticles (see **Table I**).

**Table I.** Properties of $Hf_xZr_{1-x}O_{2-y}$ nanoparticles and thin films (for comparison)

| Material | Doping | Geometry | Size/thickness (nm) | Symmetry group(s) | Ferroelectric properties | Synthesis method | Ref. |
|---|---|---|---|---|---|---|---|
| $Hf_xZr_{1-x}O_2$ | None | Particles of faceted shape | 5.5, 4.3, 3.6 | tetragonal or cubic | n/m | Sol-gel | 17 |
| $HfO_2$ | None | particle | 10 - 15 | monoclinic phase ($P2_1/c$) | n/m | auto-igniting combustion | 18 |
| $HfO_2$ | None | spherical particles | 8.79, 7.16, 6.78 | monoclinic | n/m | precipitation method | 19 |
| $HfO_2$ | None | particle | 4 - 120 | Tetragonal and monoclinic | n/m | hydrothermal route | 20 |
| $HfO_2$ | None | particle | 61 - 80 | n/m | low loss dielectric composite | US Research Nanomaterials, Inc. | 21 |
| $HfO_2$ | None | particle | 60 – 90 | monoclinic | n/m | hydrothermal synthesis | 22 |
| $HfO_2$ | Dy and Sm co-doped | particle | 10 - 31 | monoclinic and cubic phases | n/m | Pechini type sol-gel method | 23 |
| $Hf_xZr_{1-x}O_{2-y}$ | O-vacancies | particle | n/m | m<br><br>o + m | possible from XRD<br><br>Yes (theory) | Organo-nitrate; pyrogenic | 24 |
| $HfO_2$ | either $Eu^{3+}$ or $Nb^{5+}$ doping | particle | 17 - 47 | monoclinic + small amount of tetragonal phase | n/m | Sol–Gel and Combustion Synthesis | 25 |
| $Hf_{0.5}Zr_{0.5}O_2$ | None | Particles of spindle-like or spherical shape | 200 - 50<br><br>3 – 4 | ortho-rhombic $Pca2_1$ $Pbca$, $Pbcm$ | possible from XRD | Hydro-thermal synthesis from $HfCl_4$ and $ZrCl_4$ | 26 |
| $HfO_2$ | Si-, Al-, Gd-doped | film | 10 nm and 40 nm thick | Ortho-rhombic ($Pca2_1$) | polarization hysteresis in electric field | atomic layer deposition (ALD) | 27 |
| $Hf_{0.5}Zr_{0.5}O_2$ | None O-deficient (air, high-vacuum, Ar) | film | 17 nm | ortho-rhombic /tetragonal | Yes (PFM response hysteresis loop) | plasma-enhanced ALD | 28 |

This work aims to bridge this gap by investigating the charge-polarization coupling within these materials using the Landau-Ginzburg-Devonshire (LGD) model [29, 30, 31] in complex with DFT calculations, which allows to determine the "effective" Landau expansion coefficients for polar (FE) and antipolar (AFE) orderings in hafnia-based compounds. To verify the effectiveness of the LGD model, we explain the X-ray diffraction and EPR data for oxygen-deficient $HfO_{2-y}$ nanoparticles to determine their phase composition, as well as the nanoparticles prepared under varying annealing conditions (pure and doped with yttrium Y) and explain the formation of the ferroelectric-like orthorhombic phase in them.

## THEORETICAL DESCRIPTION

### A. "Effective" Landau-Ginzburg-Devonshire model

Further analysis of the spatial-temporal evolution of polarization in the nanosized $Hf_xZr_{1-x}O_{2-y}$ is conducted using a combined approach. This approach incorporates elements of the Kittel-type model [32], incorporating both polar and antipolar modes [33, 34, 35], with the theoretical framework of the Landau-Ginzburg-Devonshire (LGD) approach [29, 30, 31]. Within this framework, the free energy functional $F$ is expressed as the sum of several key terms [29 - 31]:

$$F = F_{bulk} + F_{grad} + F_{el} + F_S. \qquad (1)$$

The first term represents the bulk free energy, an expansion based on the second and fourth powers of both the polar ($P_f$) and antipolar ($A_f$) order parameters, $F_{bulk}$:

$$F_{bulk} = \int_{V_f} d^3r \left( \frac{a_P}{2} P_f^2 + \frac{b_P}{4} P_f^4 + \frac{\eta}{2} P_f^2 A_f^2 + \frac{a_A}{2} A_f^2 + \frac{b_A}{4} A_f^4 \right). \qquad (2a)$$

Here $V_f$ is the volume and surface area of the HfO$_{2-y}$ nanoparticle. Subsequent terms account for the energy contribution arising from gradients in the polarization ($F_{grad}$), the electrostatic energy ($F_{el}$), and the surface energy ($F_S$), respectively are listed in Refs.[29 - 31].

It is important to consider how the LGD expansion coefficients depend on various factors, which in turn can be influenced by temperature, size, elastic stresses and/or strains. For classical ferroelectric films with a pronounced temperature-dependent and strain-dependent soft mode, the coefficients $a_P$ and $a_A$ exhibit a linear relationship with both temperature and strain (as shown in references (see [36, 37]).

Spatial-temporal evolution of $P_f$ and $A_f$, is determined from the coupled time-dependent LGD type Euler-Lagrange equations, derived by minimizing the system's free energy $F$:

$$\Gamma_P \frac{\partial P_f}{\partial t} + a_P P_f + b_P P_f^3 + \eta A_f^2 P_f - g \Delta P_f = E_f, \qquad (3a)$$

$$\Gamma_A \frac{\partial A_f}{\partial t} + a_A A_f + b_A A_f^3 + \eta P_f^2 A_f - g \Delta A_f = 0, \qquad (3b)$$

where $\Gamma_P$ and $\Gamma_A$ are Landau-Khalatnikov relaxation coefficients, and $g$ is a positive gradient coefficient written in an isotropic approximation. The relaxation times of $P_f$ and $A_f$ are $\tau_P = \Gamma_P / |a_P|$ and $\tau_A = \Gamma_A / |a_A|$, respectively. The corresponding boundary conditions for (3) are of the third kind [38]:

$$\left( P_f + \Lambda_P \frac{\partial P_f}{\partial n} \right) \bigg|_S = 0, \quad \left( A_f + \Lambda_A \frac{\partial A_f}{\partial n} \right) \bigg|_S = 0. \qquad (4)$$

Here $\Lambda_P = \frac{g}{c_P}$ and $\Lambda_A = \frac{g}{c_A}$ are extrapolation lengths, which physical range is (0.5 – 5) nm [39]. Here $S$ is the surface of the HfO$_{2-y}$ nanoparticle.

Examples of how the effective LGD model works quantitatively are given in Refs.[29 - 31]. There polarization hysteresis loops, measured experimentally in Hf$_x$Zr$_{1-x}$O$_2$ thin films by Park et al. [9], are shown. The LGD-model parameters, determined from fitting of experimental results from Park et al. [9] in the references .[29 - 31], are listed in the last four columns of **Table II.**

**Table II.** Landau-Ginsburg-Devonshire parameters of Hf$_x$Zr$_{1-x}$O$_2$ films*

| x | phase | $P_r(x)$, C/m² | $P_{c1,2}(x)$, C/m² | $\tilde{\chi}(x)$ | $a_P(x) \times 10^9$, m/F | $a_A(x) \times 10^9$, m/F | $b_P \approx b_A \times 10^{10}$, Vm⁵/C³ | $\eta(x) \times 10^{10}$, Vm⁵/C³ |
|---|---|---|---|---|---|---|---|---|
| 0 | PE | 0 | N/A | 35 | 3.227 | N/P | N/P | N/P |
| 0.2 | AFE | 0 | 0.22, 0.17 | 40 | 0.812 | -0.936 | 0.9 | 1.93 |
| 0.3 | AFE | 0 | 0.22, 0.15 | 50 | 0.010 | -1.190 | 1.3 | 2.458 |
| 0.4 | AFE | 0 | 0.18, 0.11 | 48 | -0.104 | -0.937 | 1.1 | 2.888 |
| 0.50 | FE | 0.195 | N/A | 95 | -0.594 | N/P | 1.563 | N/P |
| 0.57 | FE | 0.173 | N/A | 76 | -0.743 | N/P | 2.483 | N/P |
| 0.70 | FE | 0.038 | N/A | 400 | -0.141 | N/P | 10.039 | N/P |
| 0.81 | PE | 0.025 | N/A | 31 | 3.643 | N/P | N/P | N/P |
| 1.00 | PE | 0 | N/A | 19 | 5.944 | N/P | N/P | N/P |
| h | films with thickness $h = 9.2$ nm | | | | | | | |

**B. Density Functional Theory calculations**

Calculations were made using density functional theory, which allows us to calculate the parameters of the ground state structure by minimizing the total energy functional. The functional is not exactly known, and the result depends on its model form. The calculations were performed using the full-electron package FPLO (full potential local orbital) for the local density approximation (LDA) and generalized gradient approximation (GGA) functionals, and the FPLO basis was used. A 12-atom hafnium oxide supercell was simulated for crystalline phases of different symmetries, with a 6×6×6 grid of Brillouin zone k-points. The equilibrium lattice parameters were found and compared with experimental data and values determined using other calculation packages. For both functionals, we confirmed that the crystal structure with monoclinic symmetry has the lowest ground state energy in the bulk HfO₂. At the same time, the difference between the energies of the polar orthorhombic phase (shown in **Fig. 1(b)**) and monoclinic phase (shown in **Fig. 1(a)**) calculated in HfO₂ structure by the LDA and GGA approximations turned out to be smaller (~30 eV) than that calculated by other calculation methods.

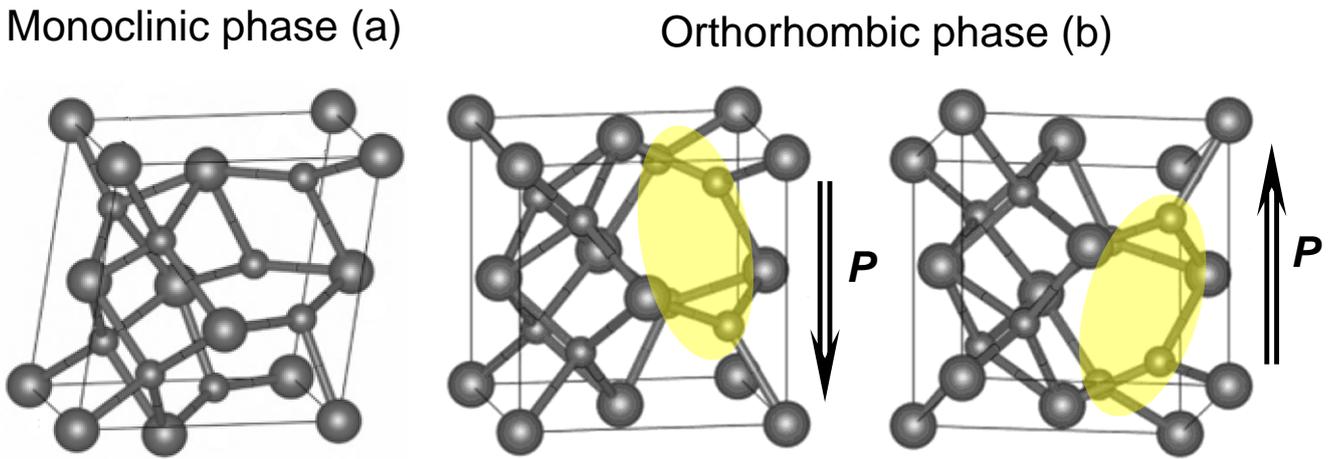

**Figure 1.** The HfO$_2$ structure with nonpolar monoclinic symmetry **(a)** and polar orthorhombic symmetry **(b)**. The polarization of the orthorhombic phase can be visualized as a shift of four oxygen atoms (shown by yellow area) along z-direction. Adapted from Ref. [40]

Thus, the DFT calculations reveal that small HfO$_2$ nanoparticles may become polar, especially in the presence of impurity atoms and/or oxygen vacancies.

## EXPERIMENTAL VERIFICATION

### A. X-ray diffraction spectra of the oxygen-deficient HfO$_{2-y}$ nanoparticles

To verify the model, we prepared several sample groups of stoichiometric HfO$_2$, oxygen-deficient HfO$_{2-y}$ and HfO$_{2-y}$ doped nanoparticles doped with Y. Namely:

- **Sample group 1** annealed at 700°C for 6 hours (or more) in air,
- **Sample group 2** annealed at 600°C for 6 hours in air + 3,8 mol% of Y$_2$O$_3$,
- **Sample group 3** annealed at 500°C for 6 hours in CO+CO$_2$.

This diverse approach yielded samples with distinct color variations, reflecting differences in their oxygen vacancy concentrations, namely for Samples from 1 to 3 color changed from white to dark grey. The average size of the nanoparticles (20 nm) was determined from TEM.

X-ray diffraction (XRD) analysis using an XRD-6000 diffractometer with Cu-K$\alpha$1 radiation (2$\theta$ = 15-70°) and the database of the International Committee for Powder Diffraction Standards (JCPDS PDF-2) was used for identification the crystallographic phases of the HfO$_2$ nanoparticles. In the result, we confirmed the coexistence of both monoclinic and ferroelectric-like orthorhombic phases inside the oxygen-deficient nanoparticles. Namely, its significant amount has been found in sample groups 2 and 3 (see **Table III**). Notably, the relative abundance of each phase exhibited a demonstrably shifting pattern based on the specific annealing protocol employed, highlighting the sensitivity of phase composition to processing parameters.

Table III. XRD studies of HfO$_{2-y}$ nanopowder sample groups

| Sample group № | Phase | Mass fraction, % | Scattering region sizes, nm | Lattice parameters | | |
|---|---|---|---|---|---|---|
| | | | | $a$, Å | $b$, Å | $c$, Å |
| 1 | m | 100 | 13 | 5,1220 | 5,1603 | 5,3025 |
| 2 | m | 16,85 | 11 | 5,1408 | 5,1771 | 5,4113 |
| | o | 83,15 | 7 | 10,1180 | 5,1353 | 5,1583 |
| 3 | m | 13,24 | 12 | 5,1250 | 5,1580 | 5,3048 |
| | o | 86,76 | 8 | 10,1089 | 5,2076 | 5,1202 |

XRD analysis revealed a progressive shift in the sample's phases with changing annealing conditions. A Sample group 1 has the pure monoclinic phase ("m") as should be for a stoichiometric HfO$_2$ nanopowder. The fraction of non-ferroelectric monoclinic phase ("m") decreased from 100% (for the sample group 1) to 13.24% (for the sample group 3), while the ferroelectric-like orthorhombic phase ("o") increased from 0% (for the sample group 1) to 86.76% (for the sample group 3), respectively. The progressive change in the phase composition, namely increasing of the ferroelectric-like orthorhombic phases, is attributed to the rise in oxygen vacancy concentration induced by the specific annealing conditions and/or rare-earth doping. Corresponding XRD spectra of oxygen-deficient HfO$_{2-y}$ nanopowders are shown in **Fig. 2** and **Fig. 3**, respectively.

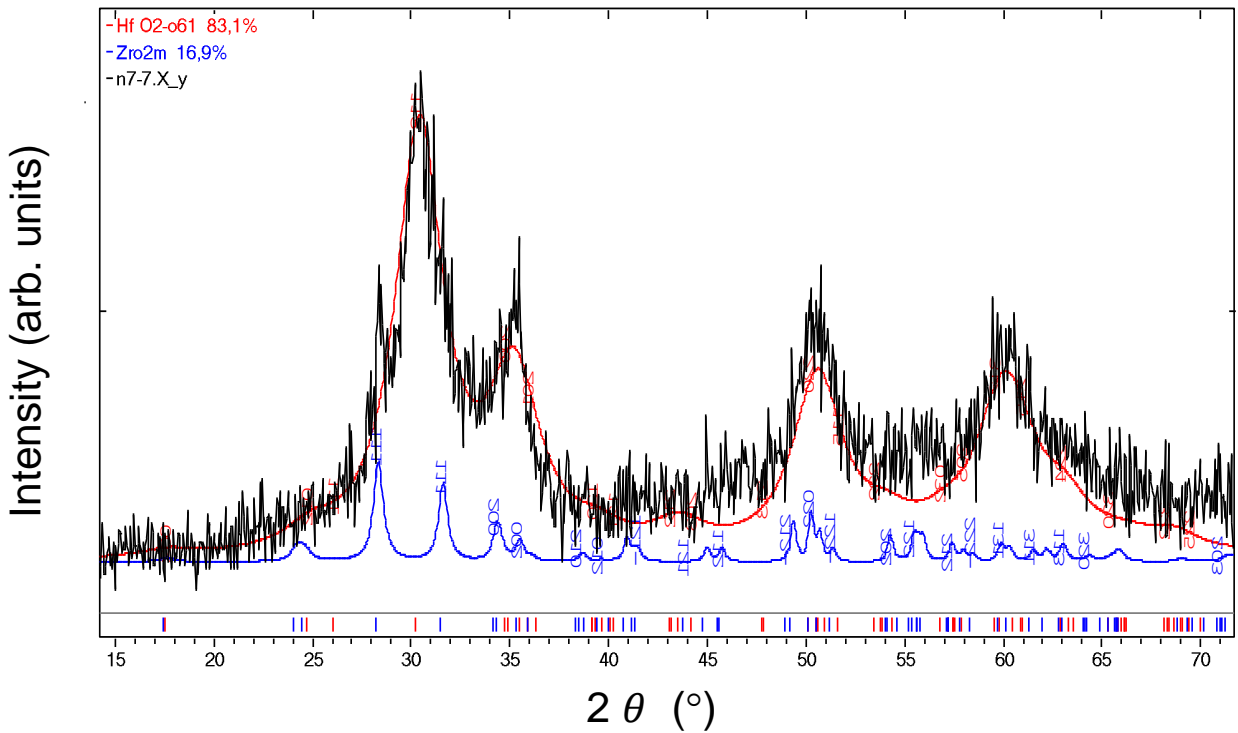

**Figure 2.** XRD spectra of oxygen-deficient HfO$_{2-y}$ nanopowders annealed at 600°C for 6 hours in air under the presence of 3,8 mol% of Y$_2$O$_3$

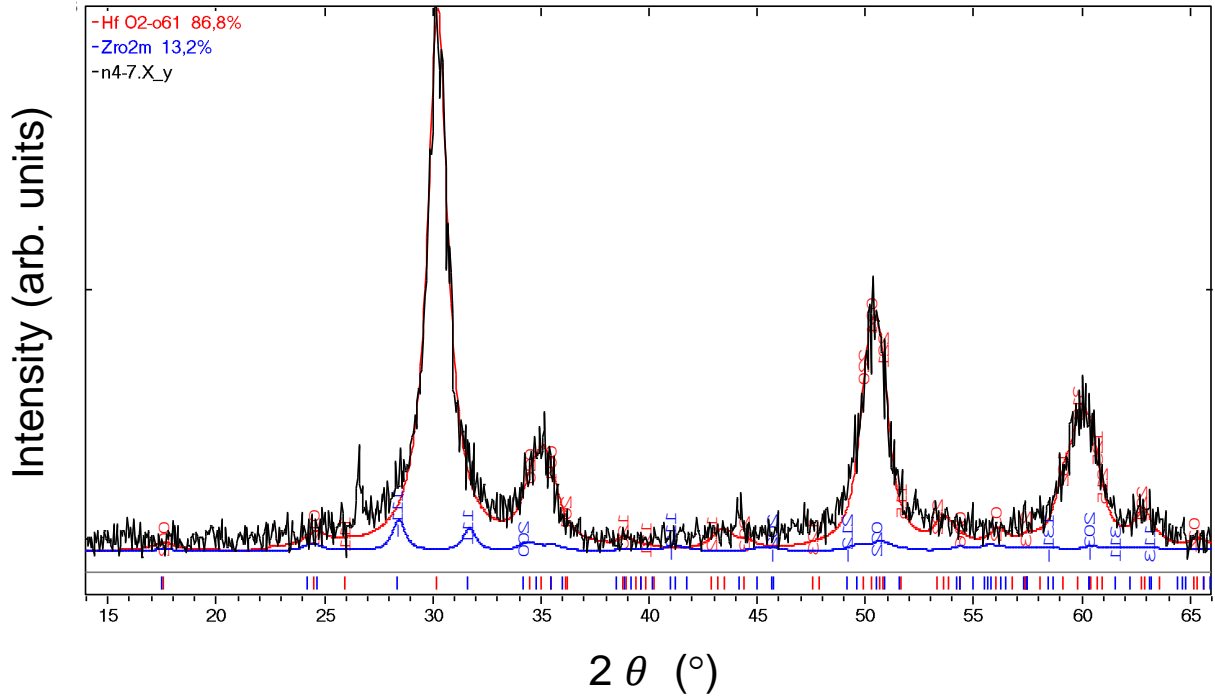

**Figure 3.** XRD spectra of oxygen-deficient $HfO_{2-y}$ nanopowders annealed at 500°C for 6 hours in $CO+CO_2$ atmosphere

This comprehensive dataset, along with the sizes of the coherent scattering regions, forms foundation for the subsequent verification of the proposed effective Landau-Ginzburg-Devonshire (LGD) model.

### B. EPR spectra the oxygen-deficient $HfO_{2-y}$ nanoparticles

EPR measurements were carried out on X-band Bruker Elexsys E580 spectrometer operating at 9.8 GHz frequency at room temperature. The EPR spectra of $HfO_{2-y}$ and $HfO_2+3,8$ mol% of $Y_2O_3$ nanoparticles are presented in **Fig. 4**. The $Y^{3+}$ and $Hf^{4+}$ ions have a closed electron shell and do not contribute to the observed EPR spectra, therefore the spectral line may be due to impurities present in the studied samples, charged oxygen vacancies or $Y^{2+}$, $Hf^{3+}$ cations that have lowered their valence state due to the formation of oxygen vacancies.

The sharp intense line in **Fig. 4(a)** belongs to organic radicals present in the sample annealed in a $CO+CO_2$ atmosphere, while the weak signal with $g_{eff} \sim 4.25$ is characteristic of $Fe^{3+}$ impurities. Much more interesting is the line with $g_{eff} \sim 1.98$ shown in the inset on **Fig. 4(a)**. This line, according to Ref.[41], can be attributed to $Hf^{3+}$ ion associated with an oxygen vacancy. The sharp, well-shaped signal apparently corresponds to Hf in the bulk of the sample, which is not affected by defects present on the surface of the nanoparticles. On the other hand, $Hf^{3+}$ located in the near-surface layer should give a broad line due to the distribution of parameters that determine the shape of the EPR line. Thus, the line with $g_{eff} \sim 2.43$ may belong to a large near-surface layer abundant in oxygen vacancies and $Hf^{3+}$ cations.

The entire spectrum shown in **Fig. 4(b)** has a very weak intensity. Its true magnitude can be estimated from the signal associated with the $Fe^{3+}$ impurity, whose intensity is close to the appropriate signal in **Fig. 4(a)**. The broad line centered at $g_{eff} \sim 2.2$ can be described as a superposition of at least two wide lines. This line can be associated with the presence of the $Y^{2+}$ ion [42] the EPR parameters of which are affected by surface defects. However, there is a significant difference from the values of the EPR parameters given in the mentioned work. Moreover, annealing in an oxygen atmosphere makes the formation of $Y^{2+}$ ions very unlikely. Thus, the nature of this line in the EPR spectrum requires further detailed study.

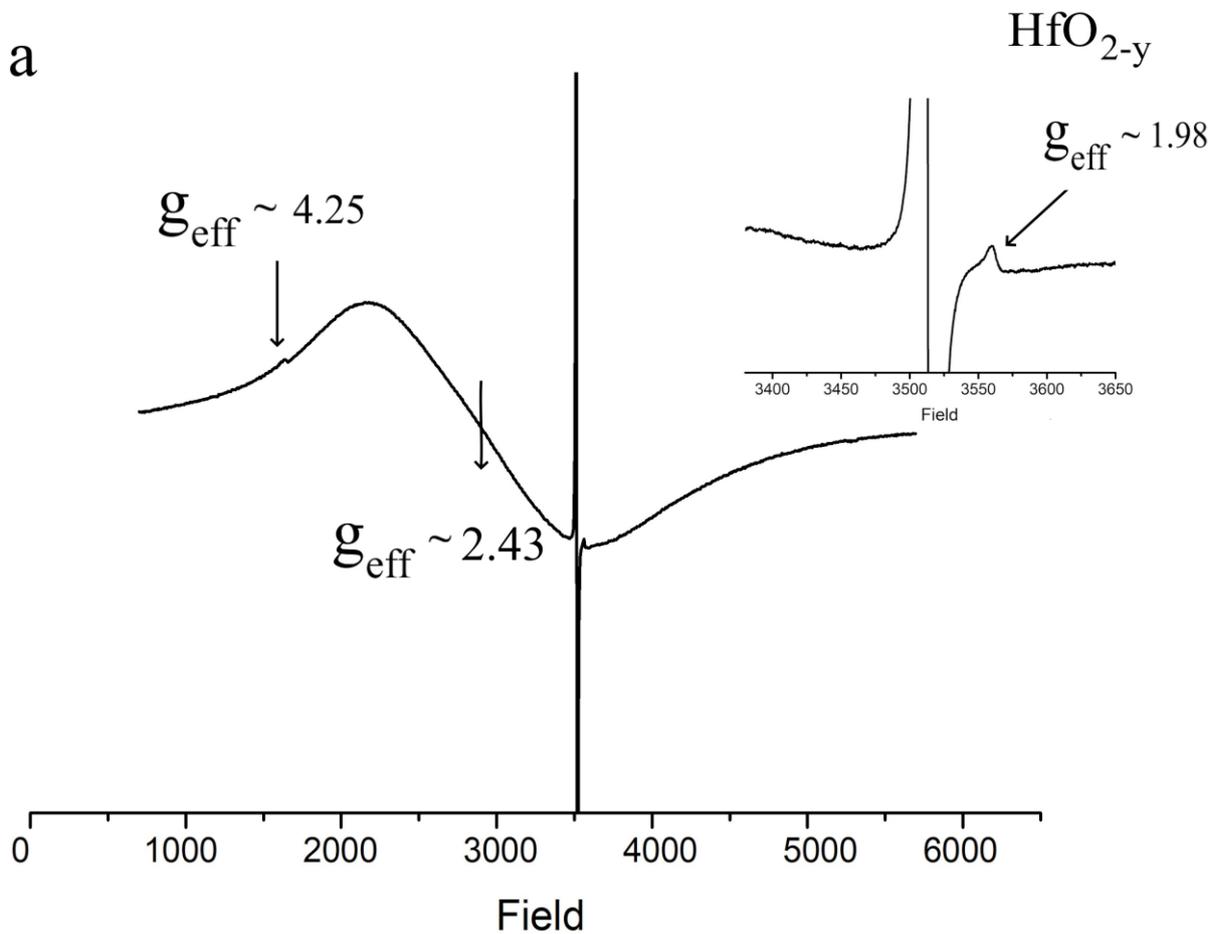

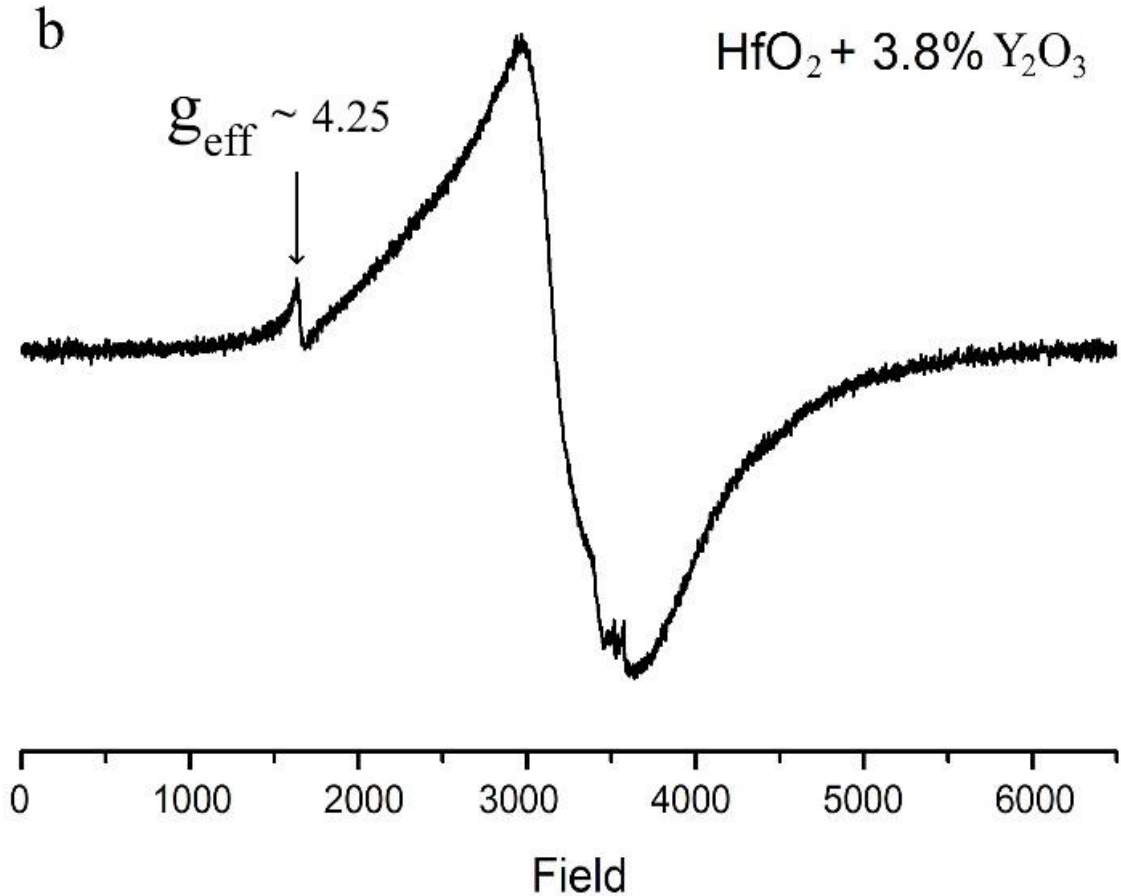

**Figure 4.** EPR spectra of oxygen-deficient $HfO_{2-y}$ nanopowders annealed at 500°C for 6 hours in $CO+CO_2$ atmosphere **(a)** and $HfO_2$+3,8 mol% of $Y_2O_3$ nanoparticles annealed at 600°C for 6 hours in air **(b)**.

## EFFECTIVE LGD MODEL FOR OXYGEN-DEFICIENT $HFO_{2-Y}$ NANOPARTICLES

Using the phase fractions and coherent scattering region sizes obtained from XRD (**Table III**), we employ the effective LGD model to simulate stoichiometric and oxygen-deficient $HfO_2$ nanoparticles. The model incorporates a vacancy concentration gradient, mimicking the presumed distribution, as shown in **Fig. 5(a)**.

These vacancies act as elastic dipoles, influencing the material's stability. The calculated polarization-field hysteresis loops (shown in **Fig. 5(b)**) reveal a progressive transition from dielectric-like to antiferroelectric and ferroelectric characteristics with increasing vacancy concentration, aligning well with the experimental XRD findings.

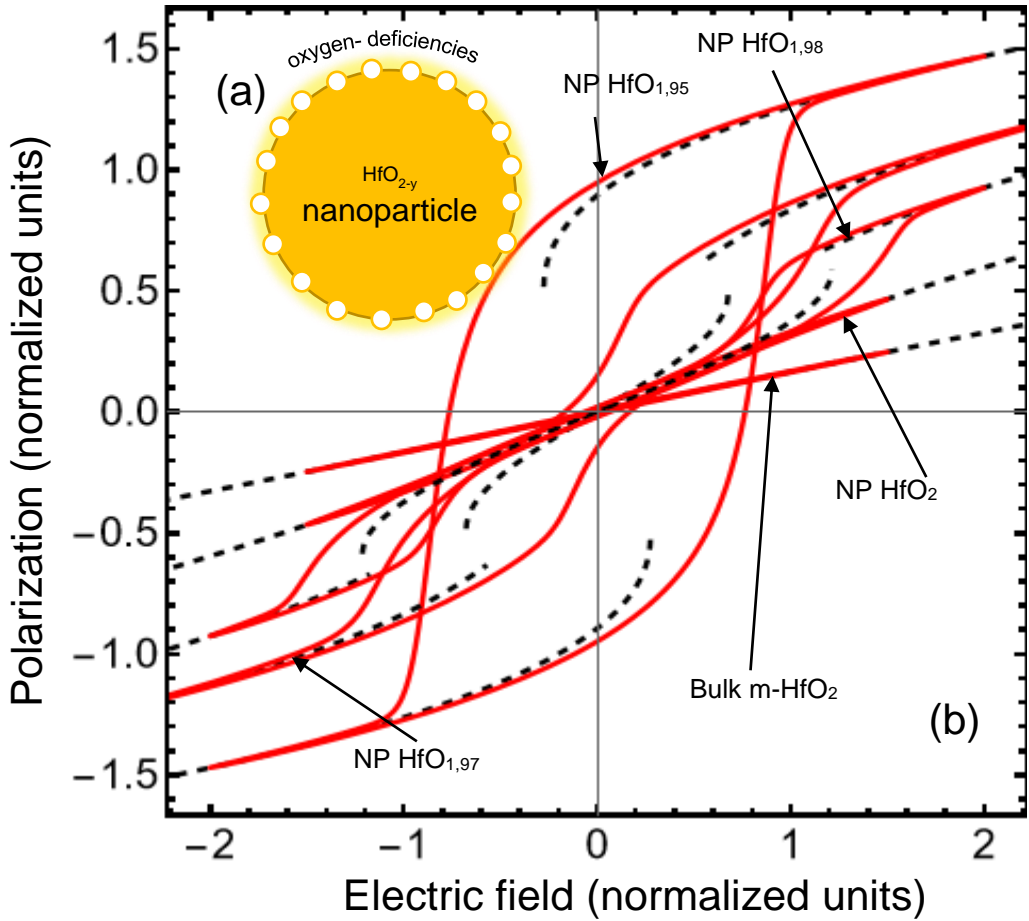

**Figure 5. (a)** The radial cross-section of the oxygen-deficient HfO$_{2-y}$ nanoparticle. **(b)** Polarization-field dependences (hysteresis loops) calculated using the effective LGD model for the stoichiometric HfO$_2$ bulk, 20-nm size stoichiometric HfO$_2$ nanoparticles (NP) and HfO$_{2-y}$ nanoparticles with different concentration of oxygen vacancies near the surface: $y = 2$ %, 3 % and 5 %. The dielectric-type, paraelectric-type, antiferroelectric-type and ferroelectric-type loops are shown. Black dotted curves show the equilibrium polarization-field dependences and red solid curves show the dynamic polarization-field hysteresis loops.

## CONCLUSIONS

Obtained results indicate the strong correlation between the crystalline structure symmetry, point defects and possible appearance of the ferroelectric-like polarization in HfO$_{2-y}$ nanoparticles. Notably, that XRD and EPR analysis revealed the formation of the ferroelectric-like orthorhombic phase in the oxygen-deficient HfO$_{2-y}$ nanoparticles (both in pure and doped with rare-earth element yttrium). Also, DFT calculations showed that small HfO$_2$ nanoparticles may become polar, especially in the presence of impurity atoms and/or oxygen vacancies.

To explain the experimental results, we have modified the effective LGD model [29-31] through a tailored parameterization approach, specifically focusing on the Landau expansion coefficients associated with polar (FE) and antipolar (AFE) orderings, which agrees with the performed DFT calculations. The effective LGD model can be useful for the development of the novel generation of silicon-compatible ferroelectric nanomaterials based on the $Hf_xZr_{1-x}O_{2-y}$.

The combined experimental and theoretical approach has potential for the field of nanotechnology and material science.

## Authors' contribution

The research idea belongs to E.A.E., A.N.M. and Y.O.Z. E.A.E. wrote the codes, performed numerical calculations and compared their results to experiments. I.V.K. performed DFT calculations. Y.O.Z. performed EPR measurements. O.V.L., V.N.P., L.P.Y., prepared the samples. M.V.K. and performed XRD measurements. A.N.M. and H.V.S. wrote the manuscript draft. All co-authors discussed the results.

## Acknowledgment

The work of E.A.E., I.V.K, Y.O.Z. and L.P.Y. are funded by the National Research Foundation of Ukraine (project "Silicon-compatible ferroelectric nanocomposites for electronics and sensors", grant N 2023.03/0127). The work of A.N.M. is supported by the Ministry of Science and Education of Ukraine (grant № PH/ 23 - 2023, "Influence of size effects on the electrophysical properties of graphene-ferroelectric nanostructures") at the expense of the external aid instrument of the European Union for the fulfillment of Ukraine's obligations in the Framework Program of the European Union for scientific research and innovation "Horizon 2020".